\documentclass[a4paper,11pt]{article}
\usepackage{pos}
\usepackage{subcaption} 
\usepackage[pscoord]{eso-pic}

\usepackage{xspace}
\newcommand{\hessj}{HESS\,J1831$-$098\xspace}
\newcommand{\psr}{PSR\,J1831$-$0952\xspace}
\newcommand{\hess}{H.E.S.S.\xspace}

\title{HESS\,J1831$-$098 - Exploring a pulsar halo scenario with \hess data}

\author*[a]{Karim Sabri}
\author[a]{Yves Gallant}
\author[a]{Justine Devin}
\author[b]{Kirsty Feijen}
\onbehalf{on behalf of the \hess collaboration}

\affiliation[a]{Laboratoire Univers et Paticules de Montpellier, Université de Montpellier, CNRS/IN2P3, Place Eugène Bataillon, 34095 Montpellier Cedex 5, France}

\affiliation[b]{Universit\'e Paris Cit\'e, CNRS, Astroparticule et Cosmologie, F-75013 Paris, France}

\emailAdd{karim.sabri@umontpellier.fr}
\emailAdd{yves.gallant@in2p3.fr}
\emailAdd{devin@lupm.in2p3.fr}
\emailAdd{feijen@apc.in2p3.fr}

\abstract{Pulsar halos are a class of extended very-high-energy (VHE) sources highlighted by the HAWC observatory towards the Geminga pulsar and PSR\,B0656$+$14. These VHE sources are interpreted as the inverse Compton emission from electrons and positrons diffusing in the interstellar medium at an inhibited rate, having escaped the pulsar wind nebula. Our aim is to search for new pulsar halos using \hess data and to constrain their physical properties.

Using a physically-motivated model of pulsar halos, we created template-based models of the spatial and energetic distributions of the expected gamma-ray emission using the Gammapy library. A promising candidate source to which this model can be effectively applied is \hessj, an extended VHE source spatially coincident with two energetic pulsars, which also exhibits spectral continuity and morphological compatibility with the ultra-high-energy source 1LHAASO\,J1831$-$1007u*. It could be powered by the radio pulsar \psr with a characteristic age of 128 kyr. We present a spectro-morphological analysis of this source with \hess data, revealing that the emission is well described with a pulsar halo model, although we cannot reject a simple 2D Gaussian morphology. We discuss the implication of the derived physical parameters of the model.}

\ConferenceLogo{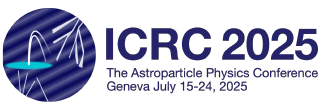}

\FullConference{39th International Cosmic Ray Conference (ICRC2025)\\
 15–24 July 2025\\
Geneva, Switzerland\\}

\begin{document}
\maketitle

\section{Introduction}
\hessj is a very-high-energy (VHE) $\gamma$-ray source first detected by \hess in 2011 with an extension of $\sim0.15^{\circ}$ \cite{sheidaei2011}. The source is coincident with a powerful middle-aged radio pulsar \psr \cite{lorimer2004} with a characteristic age $\tau_c=128~$kyr and a period $P=67~$ms ($\dot{E}=1.1\times{10^{36}}~\text{erg/s}$). More recently, a new radio pulsar PSR\,J1831$-$0941 \cite{turner2024} with $\tau_c=56$~kyr and $P=300~$ms ($\dot{E}=1.2\times{10^{35}}~\text{erg/s}$) was discovered in the vicinity of the VHE emission. Both of these pulsars could plausibly power \hessj through the creation of a pulsar wind nebula (PWN). Additionally, the 3HWC \cite{3hwc} and 1LHAASO \cite{1LHAASO} catalogs contain sources associated with the \hess source (see Discussion section).

Due to a discrepancy in its detection significance between the main and the cross-check analysis chains, \hessj was not included in the \hess Galactic Plane Survey (HGPS) catalog \cite{hgps}. We reanalyzed \hessj using the following methods: the Field of View (FoV) background method \cite{morhmann}, spectro-morphological likelihood analysis, and using templates of gas tracers to account for the large-scale Galactic diffuse emission. These methods are appropriate for extended (and possibly confused) sources near the Galactic plane.

The characteristic age of \psr is of the order 100\,kyr. According to the definition in \cite{giacinti2020} of a pulsar halo, we propose an interpretation of the VHE emission as Inverse Compton Scattering (ICS) of electrons and positrons escaping the PWN and diffusing in the interstellar medium (ISM). To that end, we adopted a frequently used model of the Geminga and Monogem (PSR\,J0633$+$1746 and PSR\,B0656$+$14) pulsar halos (and other candidates) in the \texttt{Gammapy} package for the analysis of $\gamma$-ray astronomical data \cite{gammapy:2023,gammapy1.2}.

\section{\hess data analysis}
We selected \hess observations taken between 2003 and 2016 in a $3^{\circ}$ radius around the position of \psr ($\ell=21.9^{\circ}$, $b=-0.1^{\circ}$). We excluded observations with zenith angles $\geq45^{\circ}$, and selected events within $2^{\circ}$ of the pointing direction, with an energy such that effective area is above $20\%$ of the maximum effective area, or higher than the background rate peak energy. These events are binned in an event cube (space and energy). We modeled the residual hadronic cosmic-ray (CR) background with templates built from off-source regions (using extragalactic observations), and re-normalized them with background events in each observation following the method in \cite{morhmann}. The event cubes of each observation are then stacked together by averaging the Instrument Response Function (IRF) cubes and summing the background cubes. This dataset encompasses a $6^{\circ}\times6^{\circ}$ region with a pixel size of $0.02^{\circ}$, covering reconstructed event energies from $0.5~$TeV to $100~$TeV (8 bins per decade), and true energies from $0.1~$TeV to $200~$TeV (20 bins per decade). This corresponds to $\sim45$~hours of acceptance-corrected live-time in a $0.5^{\circ}$ radius around the position of \psr.

\begin{figure}[t!]
 \centering
 \includegraphics[width=0.95\textwidth,clip]{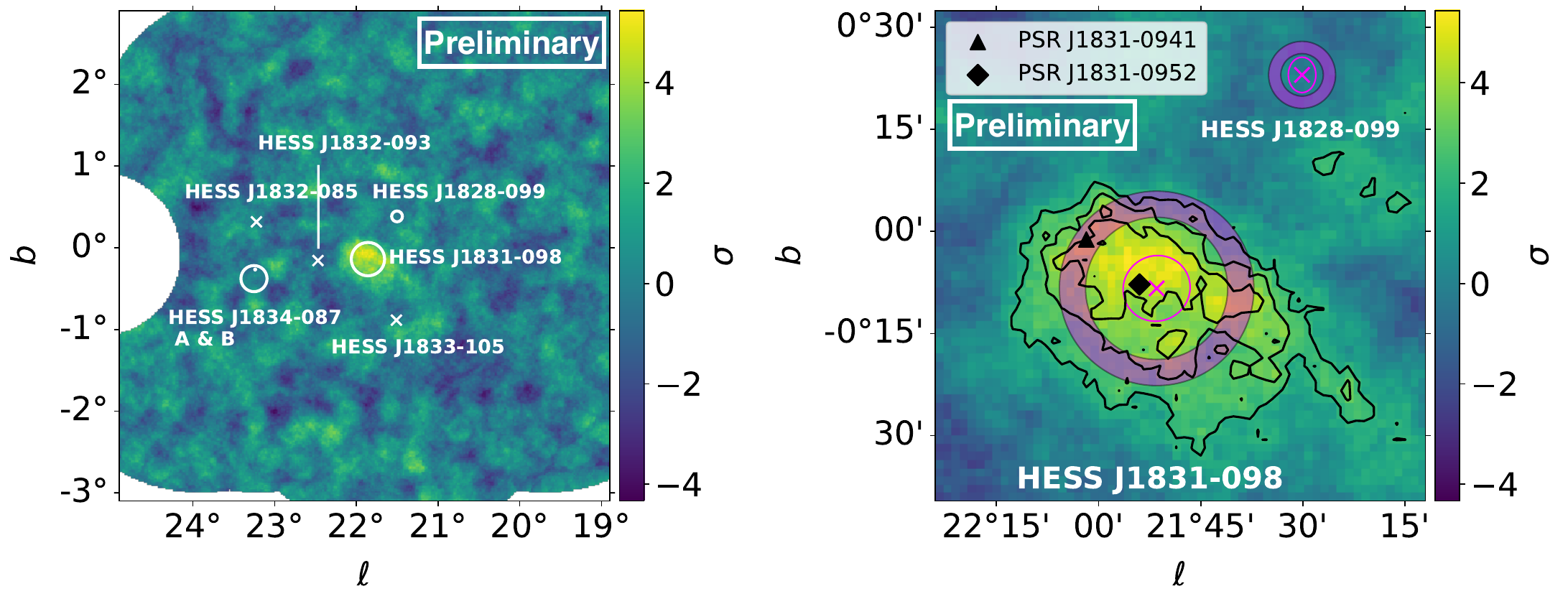} 
  \caption{\textbf{Left:} \hess significance map ($0.5-100~$TeV) with a correlation radius of $0.15^{\circ}$ after subtracting a model containing the background, diffuse emission and all HESS sources apart from \hessj. Circles represent the $1\sigma$ extents of the 2D Gaussians, and crosses represent the positions of point-like sources. The masked left-most region excludes HESS\,J1837$-$069. \textbf{Right:} Zoom of the left panel, showing the coincident pulsars. The magenta bands represent the $68\%$ confidence interval on the extensions, and the magenta circles the $95\%$ confidence interval on the positions shown with crosses. The black contours are the 2-3-4 $\sigma$ levels.}
  \label{fov_sources}
\end{figure}

We conducted a likelihood analysis by optimizing a spectro-morphological multi-component model. Assuming a constant CR density across the FoV, the large-scale Galactic diffuse emission due to pion decay traces interstellar gas. We used the Planck Dust Opacity at 353\,GHz map \cite{planck} as a proxy of the spatial gas distribution, multiplied by a spectral power law (PL) $\frac{dN}{dE}=N_0\left(\frac{E}{E_0}\right)^{-\alpha}$, with the normalization $N_0$ and the index $\alpha$ free to vary. Additionally, to account for systematic uncertainties introduced by stacking the background cubes, we introduced an energy-wise re-normalization of the residual hadronic background model during the fit. The region of interest contains 5 known sources from the HGPS catalog in addition to \hessj, which we modeled spectrally with PLs and spatially with Gaussian or point-like models as shown in Fig.~\ref{fov_sources}. The significance map (in Gaussian $\sigma$) represents the significance of the excess counts beyond the model predicted counts (with a correlation radius of $0.15^{\circ}$), following \cite{lima}.
The spatial and spectral parameters of all components were left free to fit simultaneously. The optimization of the model is done by minimizing the associated Test Statistic \cite{cash}. The significance of \hessj can be computed through hypothesis testing using Wilks' theorem \cite{wilk}.

We first modeled \hessj with a Gaussian. An alternative physically-motivated pulsar halo model involving \psr was then tested, assuming that the VHE emission is fully due to ICS of pairs diffusing in the ISM, injected by a point-like PWN. The ISM environment is described by a constant magnetic field intensity, $B=3\mu$G, and the ICS target radiation fields. We obtained the latter by fitting black-body components to the data provided in \cite{Popescu2017} at the pulsar's galactocentric position (cf. Table~\ref{tab:isrfpopescu}).

The transport in spherical coordinates of $e^{-/+}$ can be described by the continuity equation, with isotropic and homogeneous diffusion, radiative losses and a continuous point-like source term:
\begin{equation}
    \frac{\partial N(E,\mathbf{r},t)}{\partial t} - \mathbf{\nabla} [D(E) \mathbf{\nabla} N] + \frac{\partial }{\partial E}[b(E)N] = Q(E, t)\delta(\mathbf{r}-\mathbf{r_s}) \label{eq:diff_N}
\end{equation}
where $N$ is the differential particle number density at some energy $E$, time $t$ after the pulsar birth, and position $\mathbf{r}$. The diffusion coefficient  is described by$D(E)=D_0\left(\frac{E}{E_0}\right)^{\delta}$, where $D_0$ is the normalization at a reference energy $E_0$, and $\delta$ is taken as $1/3$ following Kolmogorov diffusion. $\mathbf{r_s}$ is the pulsar's position. The particle cooling rate due to synchrotron and ICS radiation is $b=-\frac{\partial E}{\partial t}$. The cooling rate due to ICS radiation is computed following \cite{Delahaye2010}. 
\begin{table}[t!]
    \centering
    \begin{tabular}{ c | c | c }
            \hline
            Optical & Near IR & Far IR \\
            \hline
            \hline
            $T=3252.76$~K & $T=546.50$~K & $T=38.92$~K \\
            $U_{rad}=2.30$~eV/cm$^3$ & $U_{rad}=0.54$~eV/cm$^3$  & $U_{rad}=1.06$~eV/cm$^3$\\
        \end{tabular}
        \caption{Radiation field parameters at PSR\,J1831$-$0952's position computed using the data from \cite{Popescu2017}.}
        \label{tab:isrfpopescu}
\end{table}

The particle injection term, $Q(E,t)$, follows a PL with exponential cutoff $E_c$ and index $\Gamma$. It is normalized such that the total energy injected in $e^{-/+}$ per unit time is equal to some fraction, $\eta$ (the injection efficiency), of the pulsar's spin-down power. We consider a minimum $e^{-/+}$ energy of $0.1~$TeV, and set $E_c=1~$PeV.

The semi-analytical solution to Eq.~\ref{eq:diff_N} can be found by defining the diffusion length scale $\lambda^2(E, E')=4\int_E^{E'} \frac{D(E'')}{b(E'')}dE''$ and the time since injection $t_{cool}=t-t'=\int_E^{E'}\frac{dE''}{b(E'')}$ \cite{syrovatskii}, with $E'$ the particle's injection energy at a time $t'$ after the pulsar's birth. We integrated from the pulsar's birth up to its true age $t_{age}=\frac{P}{(n-1)\dot{P}}\left(1-\left(\frac{P_0}{P}\right)^{n-1}\right)$ with $P$ its present-day period, $\dot{P}$ the period derivative, and $P_0$ the initial period \cite{lorimer_pulsars_book}, and we set the braking index $n=3$.

The free parameters are $D_0$, $\Gamma$, $\eta$ (which acts as a normalization) and $P_0$, which parametrizes the pulsar's true age. The position of the pulsar is fixed during the fit. For numerous combinations of the free parameter values, we computed the sky-projected $\gamma$-ray intensity using the \texttt{Naima} package \cite{naima}, assuming a distance of 3.7 kpc to \psr. We integrated over space to obtain the $\gamma$-ray Spectral Energy Distribution (SED). The normalized spatial templates are obtained by dividing the intensity by the total SED. The \texttt{Gammapy} spectral model and the spatial model are evaluated by interpolating the computed SEDs and normalized templates during the fitting procedure. The model is computed up to 200~pc away from the pulsar position.

The pulsar halo model is fitted simultaneously with the other model components. Since the pulsar halo model's parameters are highly correlated, statistical errors are estimated at a $1\sigma$ confidence level from likelihood profiles of each parameter. To assess the statistical preference of the \hessj pulsar halo model compared to a Gaussian$~\times~$PL, we used the Akaike Information Criterion \cite{aic}. For two models $M1$ and $M2$, and assuming $M2$ has a better (lower) $AIC$ score, $\Delta AIC_{M1, M2}=AIC_{M1}-AIC_{M2}=2(k_{M1}-k_{M2})+(TS_{M1}-TS_{M2})$ with $k$ the number of degrees of freedom of the corresponding model.

\section{Results}
\begin{figure}[t!]
 \centering
 \includegraphics[width=0.9\textwidth, clip]{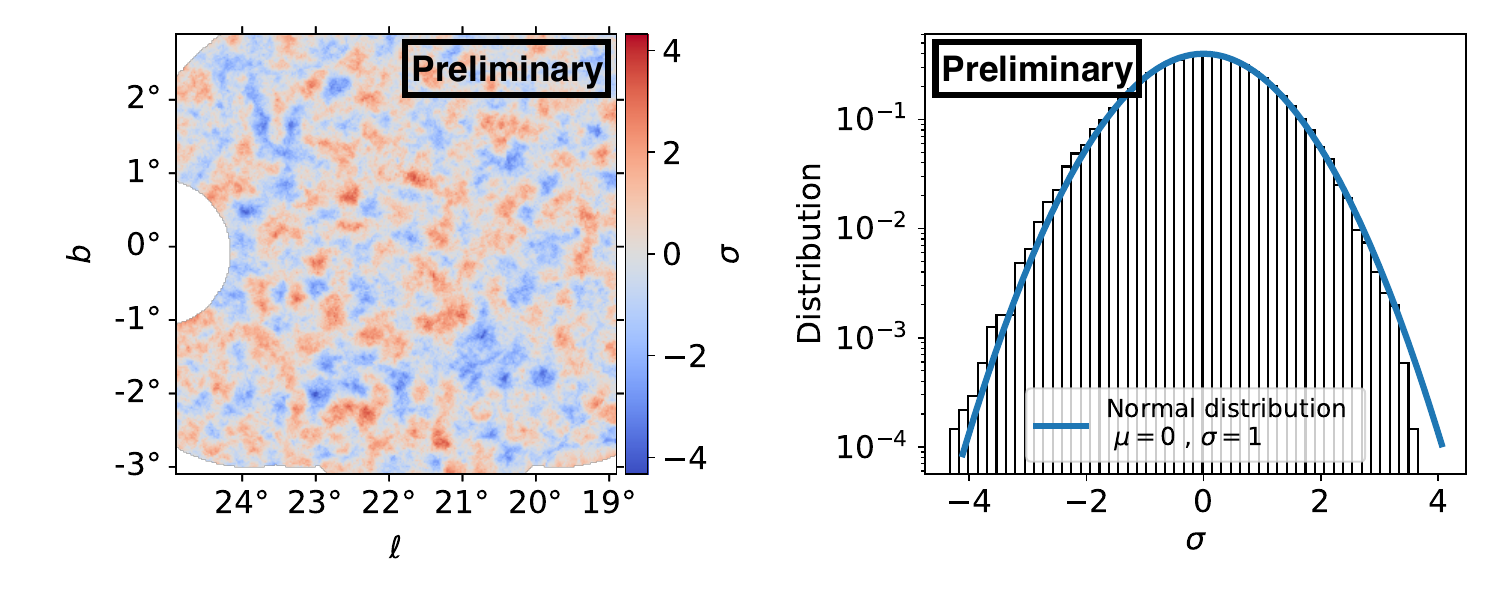}
  \caption{\textbf{Left:} Significance residual map ($0.5-100~$TeV) with a correlation radius of $0.15^{\circ}$. \textbf{Right:} Normalized distribution of the map values on the left, compared with the expectation from statistical fluctuations (i.e. a standard normal distribution).}
  \label{results_maps}
\end{figure}

We first present our results with \hessj modeled as a Gaussian$~\times~$PL. The fitted parameters for \hessj and the nearby HESS\,J1828$-$099 are shown in Table~\ref{tab:gaussresults}. The spatial residuals are shown in Fig.~\ref{results_maps}. We found a small anti-correlation between the extensions of both sources due to their proximity. We thus computed the $68\%$ confidence interval of the extensions by conducting a likelihood scan. \hessj is detected with a significance of $7.7\sigma$ (5 d.o.f.). 

\begin{figure}[t!]
 \centering
 \includegraphics[width=0.95\textwidth,clip]{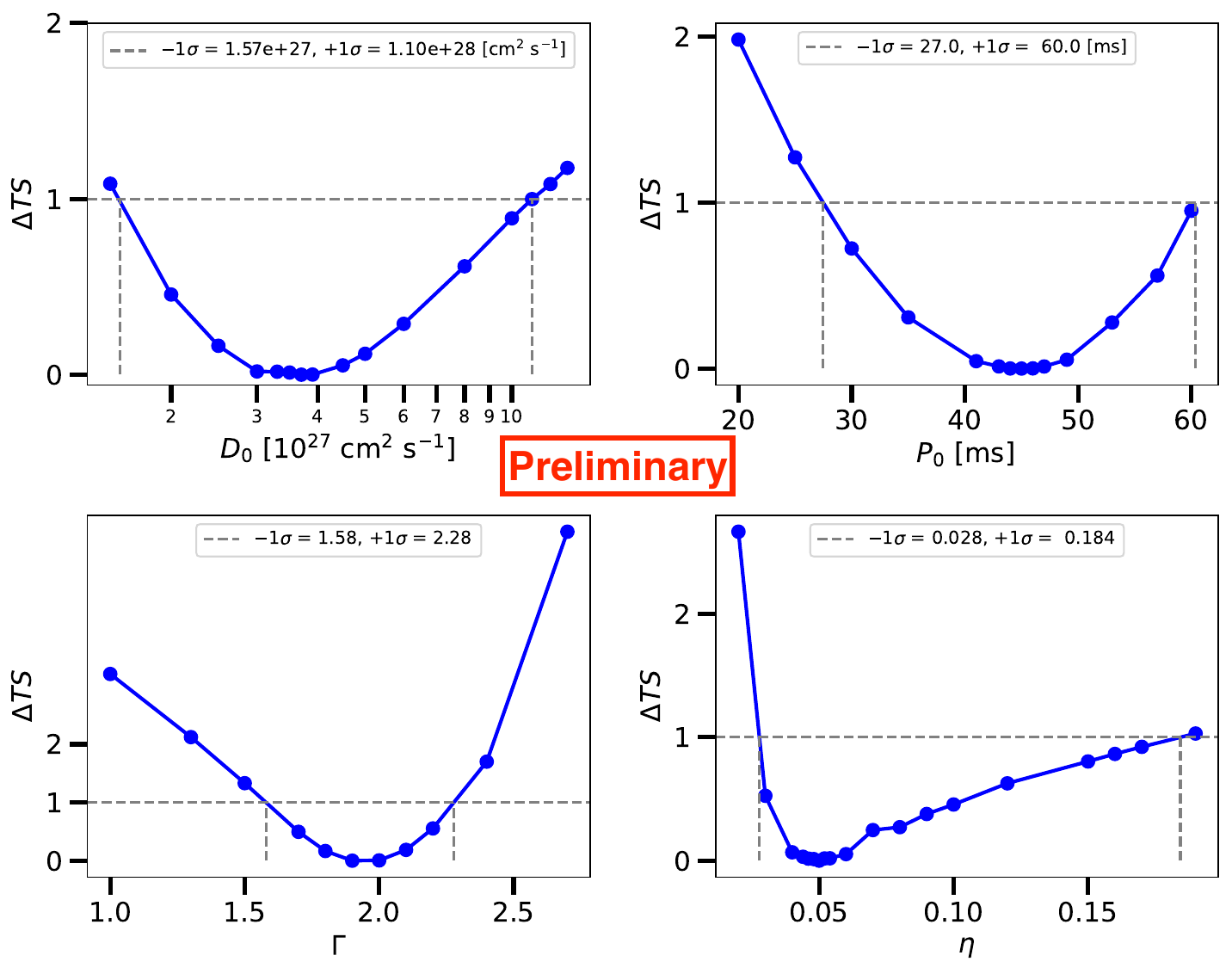} 
  \caption{Likelihood profile scans for each free pulsar halo parameter. During the scan, the parameters of all model components except the one being scanned are free.}
  \label{tsprofs}
\end{figure}

\begin{figure}[t!]
  \centering
  \hspace{-1.8cm} 
  \subfloat{\includegraphics[width=0.6\linewidth]{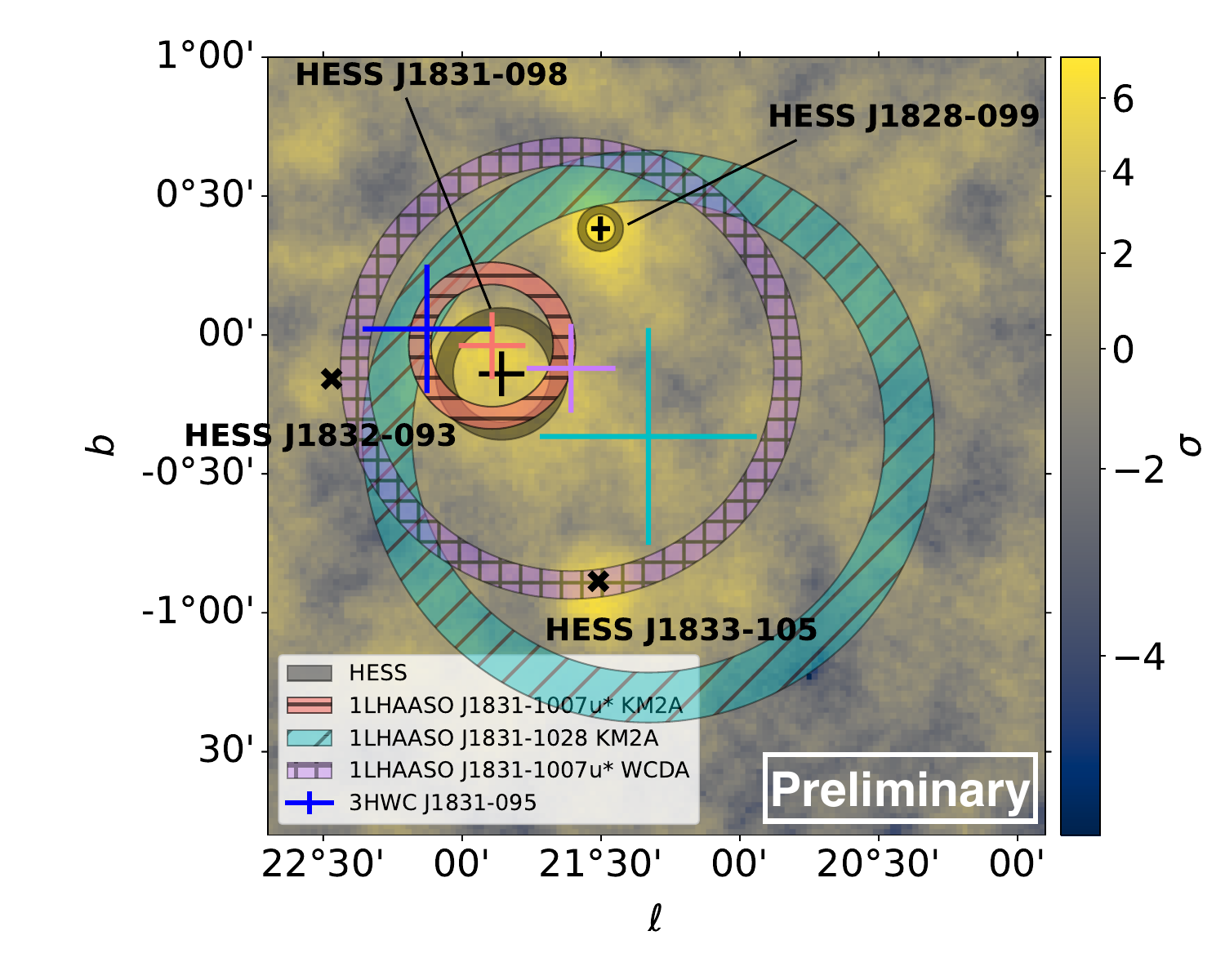}}
  \subfloat{\raisebox{1cm}{\includegraphics[width=0.5\linewidth]{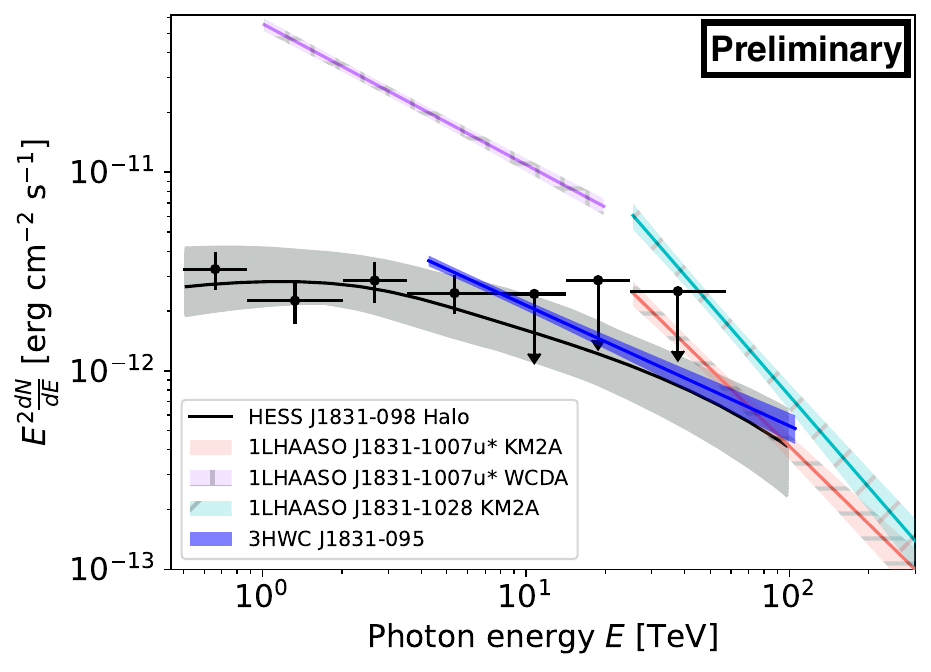}}}
\caption{\textbf{Left:} \hess significance map from $0.5-100~$TeV after subtracting a model containing the background and diffuse emission only, overlaid with our best-fit components and those reported in the 1LHAASO and 3HWC catalogs. The crosses indicate the $95\%$ confidence interval on the positions. Shaded and hatched bands represent the $68\%$ confidence interval on the Gaussian's $1\sigma$ extent. The X symbols are the \hess point-like sources' positions. \textbf{Right:} Fit photon SED and derived flux points of the pulsar halo model, the 1LHAASO components and 3HWC J1831-095. The shaded bands represent the $1\sigma$ uncertainty. The upper limits are given at the $2\sigma$ level.}
\label{spectra_and_lhaasomap}
\end{figure}

We then modeled \hessj with a pulsar halo template. The best-fit parameters are shown in Table~\ref{tab:haloresults} and are compared to those found for Geminga and Monogem by HAWC \cite{HAWCGeminga2024}. We converted the HAWC $D_0$ values to a reference energy of $100~$TeV. The likelihood profiles for each parameter are shown in Fig.~\ref{tsprofs}. The pulsar halo model (4 degrees of freedom, $M2$) has a better (lower) $AIC$ score than the Gaussian$~\times~$PL model (5 degrees of freedom, $M1$), with $\Delta AIC_{M1, M2}\sim2.1$. The derived photon SED of the pulsar halo is shown in Fig.~\ref{spectra_and_lhaasomap} and compared with those of the 1LHAASO components.
\begin{table}[t!]
    \centering
    \begin{tabular}{ c | c | c | c |c |c }
            \hline
            Component & $N_0$ & $\alpha$ & $\sigma$ [$^{\circ}$]& Lon. [$^{\circ}$] & Lat. [$^{\circ}$] \\
            \hline
            \hline
            \hessj & $1.31\pm0.15$ & $2.15\pm0.09$ & $0.20\pm0.03$ & $21.86\pm0.03$ & $-0.14\pm0.03$ \\
            \hline
            HESS\,J1828$-$099 & $0.58\pm0.08$ &  $2.35\pm0.13$ & $0.07^{+0.02}_{-0.01}$ & $21.50\pm0.01$ & $0.38\pm0.02$\\
        \end{tabular}
        \caption{Fit results for \hessj and HESS\,J1828$-$099 modeled as Gaussians. The statistical errors correspond to the $68\%$ confidence interval. $N_0$ is in units of $10^{-12}~$TeV$^{-1}$cm$^{-2}$s$^{-1}$ at a reference energy $E_0=1~$TeV.}
        \label{tab:gaussresults}
\end{table}
\begin{table}[t!]
    \centering
    \begin{tabular}{ c | c | c | c  }
            \hline
            Parameter & \hessj (this work) & Geminga \cite{HAWCGeminga2024} & Monogem \cite{HAWCGeminga2024} \\
            \hline
            \hline
            $D_0$ [$10^{27}$ cm$^{2}$s$^{-1}$] & $3.35^{+7.66}_{-1.78}$ & $4.97^{+3.38}_{-2.0}$ & $6.82^{+6.97}_{-3.39}$   \\
            \hline
            $\Gamma$ & $1.90^{+0.38}_{-0.32}$&  $0.95\pm0.32$&   $1.06\pm0.14$ \\
            \hline
            $\eta$ & $0.043^{+0.141}_{-0.015}$ & $0.066\pm0.025$ & $0.051\pm0.027$ \\
            \hline
            $P_0$ [ms] & $42.9^{+17.1}_{-15.9}$ & - &  - \\
        \end{tabular}
        \caption{Fit results for \hessj modeled as a pulsar halo, compared with HAWC's results for Geminga and Monogem \cite{HAWCGeminga2024}. The statistical errors are given as the 68\% confidence interval. For the HAWC results, we cite the systematic uncertainties.}
        \label{tab:haloresults}
\end{table}
\section{Discussion}
The $\gamma-$ray emission can be well described by a pulsar halo model, although it is not significantly preferred over a Gaussian$~\times~$PL model. The best-fit diffusion coefficient of \hessj is similar to that of the Geminga and Monogem pulsar halos. The estimated injection efficiency, $\eta$, is plausible for a pulsar-powered source. The weaker constraint on the upper bound of this parameter is due to its correlation with $P_0$. As shown in Fig.\ref{spectra_and_lhaasomap}, we only derive upper limits on the flux points beyond $\sim10~$TeV. The true age of the pulsar, which we estimated as $75.5^{+31.7}_{-50.2}~$kyr ($68\%$ C.I.), determines the contribution to the photon SED by relic $e^{-/+}$. Consequently, $P_0$ strongly impacts the photon SED at energies roughly below the energy where the particle cooling time is equal to the pulsar's true age. For the optimal value of $P_0$, this energy corresponds to $\sim7.5~$ TeV. The difference in the values of the injection index $\Gamma$ between \hessj and those of Geminga and Monogem could be due to a correlation with the injection cut-off, which is different in the HAWC analysis ($E_c=100~$TeV). 

As shown in Fig.~\ref{spectra_and_lhaasomap}, our results show good agreement between \hessj and the KM2A component of 1LHAASO\,J1831$-$1007u*. The latter is reportedly significant above $100~$TeV and shows spectral continuity and a compatible morphology with \hessj. Further, considering the systematic uncertainties discussed in \cite{3hwc} in this region, 3HWC\,J1831$-$095 could be spatially compatible with \hessj.

Finally, we computed the average energy density in $e^{-/+}$ derived from the fitted halo model. For the optimal values, the average in a volume corresponding to the fitted Gaussian's $68\%$  containment (for comparison with \cite{giacinti2020}) is $0.24~$eV~cm$^{-3}$. This value is higher than the magnetic field's energy density, however a lower injection cut-off energy could yield a lower particle energy density (assuming the same total injection efficiency).



\bibliographystyle{JHEP}
\bibliography{refs}

\end{document}